\documentstyle[epsf]{article}

\newcommand{\be}{\begin{equation}}
\newcommand{\ee}{\end{equation}}
\newcommand{\bea}{\begin{eqnarray}}
\newcommand{\eea}{\end{eqnarray}}
\newcommand{\bi}{\bibitem}
\newcommand{\nn}{\nonumber}

\newcommand{\complex}{{{\rm I} \kern -.59em {\rm C}}}
\catcode`\@=11
\def\lsim{\mathrel{\mathpalette\@versim<}}
\def\gsim{\mathrel{\mathpalette\@versim>}}
\def\@versim#1#2{\vcenter{\offinterlineskip
\ialign{$\m@th#1\hfil##\hfil$\crcr#2\crcr\sim\crcr } }}
\catcode`\@=12

\makeatletter
\let\chapter\hid@chapter
\makeatother

\begin{document}
\begin{titlepage}
  \renewcommand{\thefootnote}{\fnsymbol{footnote}}
  \begin{flushright}
    \begin{tabular}{l@{}}
      IFUNAM-FT98-2\\
      hep-ph/9803217
    \end{tabular}
  \end{flushright}

  \vskip 0pt plus 0.4fill

  \begin{center}
    {\LARGE \textbf{Gauge-Yukawa-Finite Unified Theories and their
          Predictions}\footnote{Supported partly by the mexican
          projects CONACYT 3275-PE, PAPIIT IN110296, by the EU
          projects FMBI-CT96-1212 and ERBFMRXCT960090, and by the
          Greek project PENED95/1170;1981.}\par}

  \end{center}

  \vskip 0pt plus 0.2fill

  \begin{center}
    {\large
      T.~Kobayashi%
      }\\
    \textit{
      Inst.~of Particle and Nuclear Studies,
      Tanashi, Tokyo 188, Japan
      }\\
    \vspace{1ex}
    {\large 
      J.~Kubo%
      }\\
    \textit{Dept.~of Physics, Kanazawa Univ.,
      Kanazawa 920-1192, Japan
      }\\
    \vspace{1ex}
    {\large
      M.~Mondrag\'on%
      }\\
    \textit{Inst.~de F{\'{\i}}sica, UNAM, Apdo. Postal 20-364,
      M\'exico 01000 D.F., M\'exico
      }\\
    \vspace{1ex}
    {\large
      G.~Zoupanos%
        }\\
    \textit{
      Physics Dept., Nat. Technical Univ.,
      GR-157 80 Zografou, Athens, Greece\\
    and\\
      Institut f.~Physik, Humboldt-Universit\"at, D10115 Berlin,
      Germany
      }
   
    \vskip 1ex plus 0.3fill

    {\large February 26, 1998}

    \vskip 1ex plus 0.7fill

    \textbf{Abstract}
  \end{center}
  \begin{quotation}
    Gauge-Yukawa unified theories, and in particular
  those which are finite beyond the unification scale, have been
  extended to include a soft supersymmetry breaking (SSB) sector.  In
  the case of the Finite Unified Theories a new solution to the
  condition of two-loop finiteness of the SSB parameters is found,
  which requires a sum rule for the relevant scalar masses. The sum
  rule permits violations of the universality of the scalar masses
  which is found to lead to phenomenological problems. The minimal
  supersymmetric $SU(5)$ Gauge-Yukawa model and two
  Finite-Gauge-Yukawa models have been examined using the sum rule. The
  characteristic features of these models are: a) the old agreement of
  the top quark mass prediction persists using the new data, b) the
  lightest Higgs boson is predicted to be around 120 GeV, c) the
  s-spectrum starts above 200 GeV.  
    \end{quotation}

  \vskip 0pt plus 2fill

  \setcounter{footnote}{0}

\end{titlepage}
  
\section{Introduction}

One of the most challenging problems of the Standard Model (SM) is the
{\em reduction of its free parameters}.  In a series of papers
\cite{kmz1,mondragon2,acta} it has been shown that such a reduction
can be achieved by embedding first the SM in a supersymmetric GUT, and
subsequently searching for renormalization group invariant relations
(RGI) among gauge and Yukawa couplings, which could have been
established at the Planck scale in the framework of a more fundamental
theory. Of particular interest are the Finite-Gauge-Yukawa unified
theories which apparently are touching one of the most fundamental
problems of Elementary Particle Physics. They are based on the fact
that in certain theories with vanishing one-loop $\beta$-functions
there exist RGI relations among couplings that guarantee finiteness
even to all-orders in perturbation theory. Given the searches of
strings, non-commutative geometry, and quantum groups with similar
aims, it is very interesting to note that {\em finiteness does not
  require gravity}.

\section{Reduction of Couplings and Finiteness in $N=1$ SUSY Gauge Theories}

A RGI relation among couplings,
$\Phi (g_1,\cdots,g_N) ~=~0$,
has to satisfy the partial differential equation (PDE)
$\mu\,d \Phi /d \mu ~=~
\sum_{i=1}^{N}
\,\beta_{i}\,\partial \Phi /\partial g_{i}~=~0$, 
where $\beta_i$ is the $\beta$-function of $g_i$.
There exist ($N-1$) independent  $\Phi$'s, and
finding the complete
set of these solutions is equivalent
to solve the so-called reduction equations (REs),
$\beta_{g} \,(d g_{i}/d g) =\beta_{i}~,~i=1,\cdots,N$,
where $g$ and $\beta_{g}$ are the primary
coupling and its $\beta$-function.
Using all the $(N-1)\,\Phi$'s to impose RGI relations,
one can in principle
express all the couplings in terms of
a single coupling $g$.
 The complete reduction,
which formally preserves perturbative renormalizability,
 can be achieved by
demanding a 
 power series solution,
where the uniqueness of such a power series solution
can be investigated at the one-loop level.
The completely reduced theory contains only one
independent coupling with the
corresponding $\beta$-function.
This possibility of coupling unification is
attractive, but  it can be too restrictive and hence
unrealistic. To overcome this problem, one  may use fewer $\Phi$'s
as RGI constraints.

It is clear by examining specific examples,
that the various 
couplings in supersymmetric theories have easily the same asymptotic
behaviour.  Therefore searching for a power series solution 
to the REs is justified. This is not the case in
non-supersymmetric theories.
 
Let us then consider a chiral, anomaly free,
$N=1$ globally supersymmetric
gauge theory based on a group G with gauge coupling
constant $g$. The
superpotential of the theory is given by
\bea
W&=& \frac{1}{2}\,m^{ij} \,\Phi_{i}\,\Phi_{j}+
\frac{1}{6}\,C^{ijk} \,\Phi_{i}\,\Phi_{j}\,\Phi_{k}~,
\label{supot}
\eea
where $m^{ij}$ and $C^{ijk}$ are gauge invariant tensors and
the matter field $\Phi_{i}$ transforms
according to the irreducible representation  $R_{i}$
of the gauge group $G$. 

 The one-loop $\beta$-function of the gauge
coupling $g$ is given by 
\bea
\beta^{(1)}_{g}&=&\frac{d g}{d t} =
\frac{g^3}{16\pi^2}\,[\,\sum_{i}\,l(R_{i})-3\,C_{2}(G)\,]~,
\label{betag}
\eea
where $l(R_{i})$ is the Dynkin index of $R_{i}$ and $C_{2}(G)$
 is the
quadratic Casimir of the adjoint representation of the
gauge group $G$. The $\beta$-functions of
$C^{ijk}$,
by virtue of the non-renormalization theorem, are related to the
anomalous dimension matrix $\gamma^j_i$ of the matter fields
$\Phi_{i}$ as:
\be
\beta_{C}^{ijk}=\frac{d}{dt}\,C^{ijk}=C^{ijp}\,
~\sum_{n=1}~\frac{1}{(16\pi^2)^n}\,\gamma_{p}^{k(n)} +(k
\leftrightarrow i) +(k\leftrightarrow j)~.
\label{betay}
\ee
At one-loop level $\gamma^j_i$ is 
\be
\gamma_i^{j(1)}=\frac{1}{2}C_{ipq}\,C^{jpq}-2\,g^2\,C_{2}(R_{i})\delta_i^j~,
\label{gamay}
\ee
where $C_{2}(R_{i})$ is the quadratic Casimir of the representation
$R_{i}$, and $C^{ijk}=C_{ijk}^{*}$.

As one can see from Eqs.~(\ref{betag}) and (\ref{gamay}) 
 all the one-loop $\beta$-functions of the theory vanish if
 $\beta_g^{(1)}$ and $\gamma_i^{j(1)}$ vanish, i.e.
\be
\sum _i \ell (R_i) = 3 C_2(G) \,,~~~~~~~~
\frac{1}{2}C_{ipq} C^{jpq} = 2\delta _i^j g^2  C_2(R_i)\,.
\label{2nd}
\ee

A very interesting result is that the conditions (\ref{2nd}) are
necessary and sufficient for finiteness at
the two-loop level.

The one- and two-loop finiteness conditions (\ref{2nd}) restrict
considerably the possible choices of the irreps.~$R_i$ for a given
group $G$ as well as the Yukawa couplings in the superpotential
(\ref{supot}).  Note in particular that the finiteness conditions cannot be
applied to the supersymmetric standard model (SSM), since the presence
of a $U(1)$ gauge group is incompatible with the condition
(\ref{2nd}), due to $C_2[U(1)]=0$.  This naturally leads to the
expectation that finiteness should be attained at the grand unified
level only, the SSM being just the corresponding, low-energy,
effective theory.

A natural question to ask is what happens at higher loop orders.
There exists a very interesting theorem \cite{LPS} which guarantees
the vanishing of the $\beta$-functions to all orders in perturbation
theory, if we demand reduction of couplings, and that all the one-loop
anomalous dimensions of the matter field in the completely and
uniquely reduced theory vanish identically.

\section{Supersymmetry breaking terms}

The above described method of reducing the dimensionless couplings has
been extended \cite{kmz-plb389} in the soft supersymmetry breaking
(SSB) dimensionfull parameters of $N=1$ supersymmetric theories. In
addition it was found \cite{kkk-plb705} that RGI SSB scalar masses in
Gauge-Yukawa unified models satisfy a universal sum rule.
Here we would like to describe how the use of the available two-loop RG
functions and the requirement of finiteness of the SSB parameters up to
this order leads to the soft scalar-mass sum rule \cite{kkmz-npb}.

Consider the superpotential given by (\ref{supot}) 
along with the Lagrangian for SSB terms,
\bea
-{\cal L}_{\rm SB} &=&
\frac{1}{6} \,h^{ijk}\,\phi_i \phi_j \phi_k
+
\frac{1}{2} \,b^{ij}\,\phi_i \phi_j
+
\frac{1}{2} \,(m^2)^{j}_{i}\,\phi^{*\,i} \phi_j+
\frac{1}{2} \,M\,\lambda \lambda+\mbox{H.c.}~
\eea
where the $\phi_i$ are the
scalar parts of the chiral superfields $\Phi_i$ , $\lambda$ are the gauginos
and $M$ their unified mass.
Since we would like to discuss 
only finite theories here, we assume that 
the gauge group is  a simple group and the one-loop
$\beta$ function of the 
gauge coupling $g$  vanishes.
We also assume that the reduction equations
admit power series solutions of the form
\bea 
C^{ijk} &=& g\,\sum_{n=0}\,\rho^{ijk}_{(n)} g^{2n}~,
\label{Yg}
\eea 
According to the finiteness theorem
of ref.~\cite{LPS}, the theory is then finite to all orders in
perturbation theory, if, among others, the one-loop anomalous dimensions
$\gamma_{i}^{j(1)}$ vanish.  The one- and two-loop finiteness for
$h^{ijk}$ can be achieved by 
\bea h^{ijk} &=& -M C^{ijk}+\dots =-M
\rho^{ijk}_{(0)}\,g+O(g^5)~.
\label{hY}
\eea

Now, to obtain the two-loop sum rule for 
soft scalar masses, we assume that 
the lowest order coefficients $\rho^{ijk}_{(0)}$ 
and also $(m^2)^{i}_{j}$ satisfy the diagonality relations
\bea
\rho_{ipq(0)}\rho^{jpq}_{(0)} &\propto & \delta_{i}^{j}~\mbox{for all} 
~p ~\mbox{and}~q~~\mbox{and}~~
(m^2)^{i}_{j}= m^{2}_{j}\delta^{i}_{j}~,
\label{cond1}
\eea
respectively.
Then we find the following soft scalar-mass sum
rule
\bea
(~m_{i}^{2}+m_{j}^{2}+m_{k}^{2}~)/
M M^{\dag} &=&
1+\frac{g^2}{16 \pi^2}\,\Delta^{(1)}+O(g^4)~
\label{sumr} 
\eea
for i, j, k with $\rho^{ijk}_{(0)} \neq 0$, where $\Delta^{(1)}$ is
the two-loop correction, which vanishes for the
universal choice in accordance with the previous findings of
ref.~\cite{jack3}.
 
\section{Gauge-Yukawa-Unified Theories}
\subsection{Finite Unified Models}

A predictive Gauge-Yukawa unified $SU(5)$ model which is finite to all
orders, in addition to the requirements mentioned already, should also
have the following properties:

\begin{enumerate}

\item 
One-loop anomalous dimensions are diagonal,
i.e.,  $\gamma_{i}^{(1)\,j} \propto \delta^{j}_{i} $,
according to the assumption (\ref{cond1}).

\item
Three fermion generations, $\overline{\bf 5}_{i}~~
(i=1,2,3)$, obviously should not couple to ${\bf 24}$.
This can be achieved for instance by imposing $B-L$ 
conservation.

\item
The two Higgs doublets of the MSSM should mostly be made out of a 
pair of Higgs quintet and anti-quintet, which couple to the third
generation.
\end{enumerate}

In the following we discuss two versions of the all-order finite
model.
\newline
${\bf A}$:  The model of ref. \cite{kmz1}.
\newline
${\bf B}$:  A slight variation of  the 
 model ${\bf A}$.

The  superpotential which describe the two models 
takes the form \cite{kmz1,kkmz-npb}
\bea
W &=& \sum_{i=1}^{3}\,[~\frac{1}{2}g_{i}^{u}
\,{\bf 10}_i{\bf 10}_i H_{i}+
g_{i}^{d}\,{\bf 10}_i \overline{\bf 5}_{i}\,
\overline{H}_{i}~] +
g_{23}^{u}\,{\bf 10}_2{\bf 10}_3 H_{4} \\
 & &+g_{23}^{d}\,{\bf 10}_2 \overline{\bf 5}_{3}\,
\overline{H}_{4}+
g_{32}^{d}\,{\bf 10}_3 \overline{\bf 5}_{2}\,
\overline{H}_{4}+
\sum_{a=1}^{4}g_{a}^{f}\,H_{a}\, 
{\bf 24}\,\overline{H}_{a}+
\frac{g^{\lambda}}{3}\,({\bf 24})^3~,\nn
\label{super}
\eea
where 
$H_{a}$ and $\overline{H}_{a}~~(a=1,\dots,4)$
stand for the Higgs quintets and anti-quintets.

The non-degenerate and isolated solutions to $\gamma^{(1)}_{i}=0$ for
the models $\{ {\bf A}~,~{\bf B} \}$ are: 
\bea (g_{1}^{u})^2
&=&\{\frac{8}{5},\frac{8}{5} \}g^2~, ~(g_{1}^{d})^2
=\{\frac{6}{5},\frac{6}{5}\}g^2~,~
(g_{2}^{u})^2=(g_{3}^{u})^2=\{\frac{8}{5},\frac{4}{5}\}g^2~,\label{SOL5}\\
(g_{2}^{d})^2 &=&(g_{3}^{d})^2=\{\frac{6}{5},\frac{3}{5}\}g^2~,~
(g_{23}^{u})^2 =\{0,\frac{4}{5}\}g^2~,~
(g_{23}^{d})^2=(g_{32}^{d})^2=\{0,\frac{3}{5}\}g^2~,
\nn\\
(g^{\lambda})^2 &=&\frac{15}{7}g^2~,~ (g_{2}^{f})^2
=(g_{3}^{f})^2=\{0,\frac{1}{2}\}g^2~,~ (g_{1}^{f})^2=0~,~
(g_{4}^{f})^2=\{1,0\}g^2~.\nn 
\eea 
According to the theorem of
ref.~\cite{LPS} these models are finite to all orders.  After the
reduction of couplings the symmetry of $W$ is enhanced
\cite{kmz1,kkmz-npb}.

The main difference of the models
${\bf A}$ and ${\bf B}$ is
that three pairs of Higgs quintets and anti-quintets couple to 
the ${\bf 24}$ for ${\bf B}$ so that it is not necessary 
to mix
them with $H_{4}$ and $\overline{H}_{4}$ in order to
achieve the triplet-doublet splitting after the symmetry breaking 
of $SU(5)$.

\subsection{ The minimal supersymmetric $SU(5)$ model}

Next let us consider the minimal supersymmetric $SU(5)$ model.  The
field content is minimal. Neglecting the CKM mixing, one starts with
six Yukawa and two Higgs couplings. We then require GYU to occur among
the Yukawa couplings of the third generation and the gauge coupling.
We also require the theory to be completely asymptotically free.  In
the one-loop approximation, the GYU yields $g_{t,b}^{2}
~=~\sum_{m,n=1}^{\infty} \kappa^{(m,n)}_{t,b}~h^m\,f^n~g^2$ ($h$ and
$f$ are related to the Higgs couplings). Where $h$ is allowed to vary
from $0$ to $15/7$, while $f$ may vary from $0$ to a maximum which
depends on $h$ and vanishes at $h=15/7$.  As a result, it was obtained
\cite{mondragon2}: $ 0.97\,g^2 \lsim g_{t}^{2} \lsim 1.37\,g^2~,~
~0.57\,g^2 \lsim g_{b}^{2}=g_{\tau}^{2} \lsim 0.97\,g^2$.  It was
found \cite{mondragon2,acta} that consistency with proton decay
requires $g_t^2,~g_b^2$ to be very close to the left hand side values
in the inequalities.

\section{Predictions of Low Energy Parameters}
 
Since the gauge symmetry is spontaneously broken below $M_{\rm GUT}$,
the finiteness conditions do not restrict the renormalization property
at low energies, and all it remains are boundary conditions on the
gauge and Yukawa couplings (\ref{SOL5}) and the $h=-MC$ relation
(\ref{hY}) and the soft scalar-mass sum rule (\ref{sumr}) at $M_{\rm
  GUT}$.  So to find the predictions of these models we examine the
evolution of these parameters according to their renormalization group
equations at two-loop for dimensionless parameters and at one-loop for
dimensional ones with these boundary conditions.  Below $M_{\rm GUT}$
their evolution is assumed to be governed by the MSSM. We further
assume a unique supersymmetry breaking scale $M_{s}$ so that below
$M_{s}$ the SM is the correct effective theory.

The predictions for the top quark mass $M_t$ are $\sim 183$ and $\sim
174$ GeV in models $\bf A$ and $\bf B$ respectively, and 
$\sim 181$ GeV for the minimal $SU(5)$ model. Comparing these
predictions with the most recent experimental value $ M_t = (175.6 \pm
5.5)$ GeV, and recalling that the theoretical values for $M_t$ may
suffer from a correction of less than $\sim 4 \%$ \cite{acta}, we see
that they are consistent with the experimental data.

Turning now to the SSB sector of these models we first note that
preliminary results on the minimal $SU(5)$ model show that the low
energy SSB sector contains a single arbitrary parameter, the gaugino
mass M, which characterizes the scale of supersymmetry breaking. The
lightest supersymmetric particle is found to be a neutralino of $\sim
220$ GeV for $M(M_{GUT})\sim 0.5$ TeV.  Concerning the SSB sector of the
finite theories $\bf A$, $\bf B$ we look for the parameter space in
which the lighter s-tau mass squared $m^2_{\tilde \tau}$ is larger
than the lightest neutralino mass squared $m^2_\chi$ (which is the
LSP).  For the case where all the soft scalar masses are universal at
the unfication scale, there is no region of $M_s=M$ below $O$(few) TeV
in which $m^2_{\tilde \tau} > m^2_\chi$ is satisfied.  But once the
universality condition is relaxed this problem can be solved naturally
(provided the sum rule). More specifically, using the sum rule
(\ref{sumr}) and imposing the conditions a) successful radiative
electroweak symmetry breaking b) $m_{\tilde\tau^2}>0$ and c)
$m_{\tilde\tau^2}> m_{\chi^2}$, we find a comfortable parameter space
for both models (although model $\bf B$ requires large $M\sim 1$ TeV).
The particle spectrum of models $\bf A$ and $\bf B$ in turn is
calculated in terms of 3 and 2 parameters respectively.

In Fig.~1 we present  the $m_{\bf 10}$ dependence of 
$m_h$ for for $M= 0.8$ (dashed) $1.0$ (solid) TeV
for the finite Model $\bf B$, which shows that the value of $m_h$ is
stable. Similar results hold also for
the minimal supersymmetric $SU(5)$ model. In particular, in Fig.2 it is shown
the dependence of the lightest Higgs mass $m_h$ in terms of the single free
parameter $M$.

\begin{figure}
           \epsfxsize= 6 cm   %or \epsfysize= HEIGHT cm
           \centerline{\epsffile{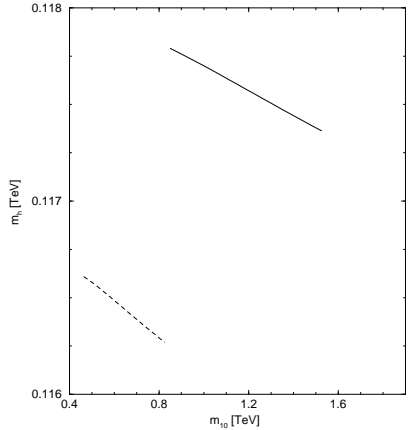}}
           \vspace*{-1cm}
        \caption{ $m_h$ as function of 
$m_{\bf 10}$ for $M= 0.8$ (dashed) $1.0$ (solid) TeV.}
        \label{fig:2}
        \end{figure}

\begin{figure}
           \epsfxsize= 6 cm   %or \epsfysize= HEIGHT cm
           \centerline{\epsffile{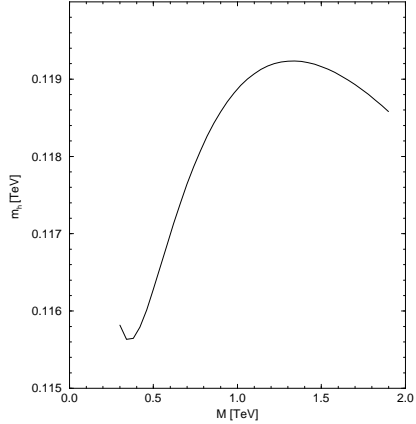}}
           \vspace*{-1cm}
        \caption{The $M$ dependence of $m_h$ for the minimal $SU(5)$
          model.} 
        \label{fig:1}
        \end{figure}

\section{Conclusions}

The search for realistic Finite Unified theories started a few years
ago \cite{kmz1,mondragon2,acta} with the successful prediction of the
top quark mass, and it has now been complemented with a new important
ingredient concerning the finiteness of the SSB sector of the theory.
Specifically, a sum rule for the soft scalar masses has been obtained
which quarantees the finiteness of the SSB parameters up to two-loops
\cite{kkmz-npb}, avoiding at the same time serious phenomenological
problems related to the previously known ``universal'' solution.  It
is found that this sum rule coincides with that of a certain class of
string models in which the massive string modes are organized into
$N=4$ supermultiplets.  Using the sum rule we can now determine the
spectrum of realistic models in terms of just a few parameters. In
addition to the successful prediction of the top quark mass the
characteristic features of the spectrum are that 1) the lightest Higgs
mass is predicted $\sim 120$ GeV and 2) the s-spectrum starts above
200 GeV. Therefore, the next important test of these Finite Unified
theories will be given with the measurement of the Higgs mass, for
which the models show an appreciable stability in their prediction.
The minimal supersymmetric $SU(5)$ gives similar results.


\begin{thebibliography}{100}

\bibitem{kmz1} D. Kapetanakis, M. Mondrag{\' o}n and
    G. Zoupanos, {\sl Zeit. f. Phys.} {\bf C60} (1993) 181;
    M. Mondrag{\' o}n and G. Zoupanos, {\sl Nucl.~Phys.~}{\bf B}
    (Proc.~Suppl) {\bf 37C} (1995) 98.

\bi{mondragon2} J. Kubo, M. Mondrag\'on and G. Zoupanos,
{\sl Nucl.~Phys.~}{\bf B424} (1994) 291.

\bi{LPS} C. Lucchesi, O. Piguet and K. Sibold,
                {\sl Helv. Phys. Acta} {\bf 61} (1988) 321; {\sl
                Phys.~Lett.~}{\bf B201} (1988) 241.

\bi{jack3}I. Jack and D.R.T. Jones,  Phys.~Lett.~{\bf B333} (1994) 372.

\bi{acta} For an extended discussion and a complete list of
references see: J.~Kubo, M.~Mondrag\'on and G.~Zoupanos, {\sl Acta
Phys.~Polon.~}{\bf B27} (1997) 3911.

\bi{kmz-plb389} J.~Kubo, M.~Mondrag\'on and G.~Zoupanos, {\sl
  Phys.~Lett.~}{\bf B389} (1996) 523.

\bi{kkk-plb705} Y.~Kawamura, T.~Kobayashi and J.~Kubo, {\sl
  Phys.~Lett.~}{\bf B705} (1997) 64.

\bi{kkmz-npb} T.~Kobayashi, J.~Kubo, M.~Mondrag\'on and G.~Zoupanos,
{\em Constraints on Finite Soft SUSY Breaking Terms} to be published
in Nucl.~Phys.~B.

\end{thebibliography}
\end{document}